\documentclass{aastex6}
\newcommand{\kms}{km s$^{-1}\;$}
\newcommand{\kmss}{km s$^{-1}$}

\newcommand{\vlsr}{V$_{\rm LSR}$}

\newcommand{\sm}{\mbox{M$_{\sun}\;$}}
\newcommand{\lsun}{\mbox{L$_{\sun}$}}

\newcommand{\ho}{H$_{2}$O$\;$}

\newcommand{\mb}{mJy beam$^{-1}$}

\newcommand{\nrest}{$\nu_{\rm {rest}}$}
\begin{document}



\title{A SEARCH FOR SUB-MILLIMETER H$_2$O MASERS IN ACTIVE GALAXIES  \\
    --  THE DETECTION OF 321 GHz H$_2$O MASER EMISSION IN NGC\,4945}
\author{Yoshiaki Hagiwara\altaffilmark{1}, Shinji Horiuchi\altaffilmark{2}, Akihiro Doi\altaffilmark{3}, Makoto Miyoshi\altaffilmark{4}, and Philip G. Edwards\altaffilmark{5}}
%
\altaffiltext{1}{Natural Science Laboratory, Toyo University, 5-28-20, Hakusan, Bunkyo-ku, Tokyo 112-8606, Japan; yhagiwara@toyo.jp}
\altaffiltext{2}{CSIRO Astronomy and Space Science, Canberra Deep Space Communications Complex,
PO Box 1035, Tuggeranong, ACT 2901, Australia}
\altaffiltext{3}{The Institute of Space and Astronautical Science, Japan Aerospace Exploration Agency, 3-1-1, Yoshinodai, Chuou-ku,  Sagamihara, Kanagawa 252-5210, Japan}
\altaffiltext{4}{National Astronomical Observatory of Japan, 2-21-1, Osawa, Mitaka, 181-8588 Tokyo, Japan}
\altaffiltext{5}{CSIRO Astronomy and Space Science, PO Box 76, Epping NSW 1710, Australia}








\begin{abstract} 
We present further results of a search for extragalactic submillimeter \ho masers using the Atacama Large Millimeter/Submillimeter Array (ALMA). The detection of a 321\,GHz \ho maser in the nearby Type 2 Seyfert galaxy, the Circinus galaxy, has previously been reported, and here the spectral analysis of four other galaxies is described. {We have discovered \ho maser emission at 321\,GHz} toward the center of NGC\,4945, a nearby Type 2 Seyfert.
The maser {emission} shows Doppler-shifted velocity features with velocity ranges similar to those of the {previously reported 22\,GHz \ho masers}, however the non-contemporaneous observations also show differences in velocity offsets. The sub--parsec-scale distribution of the 22\,GHz \ho masers revealed by earlier VLBI (Very Long Baseline Interferometry) observations suggests that the submillimeter masers could arise in an edge-on rotating disk.
The maser features remain unresolved by the synthesized beam of $\sim$ $0\arcsec$.54 ($\sim$30 pc) and are located toward the 321\,GHz continuum peak within errors. A marginally detected  (3 $\sigma$) high-velocity feature is redshifted by 579 \kms with respect to the systemic velocity of the galaxy. Assuming that this feature is real and arises from a Keplerian rotating disk in this galaxy, it is located at a radius of $\sim$0.020 pc ($\sim$1.5 $\times$ 10$^5$ Schwarzschild radii), {which would enable molecular material closer to the central engine to be probed than the 22 GHz \ho masers.}
This detection confirms that submillimeter \ho masers are a potential tracer of the circumnuclear regions of active galaxies, which will benefit from higher angular resolution studies with ALMA. 
\end{abstract}
%

\section{INTRODUCTION} 

{Maser emission at the \ho 6$_{16}$--5$_{23}$ transition (rest frequency, \nrest = 22.23508\,GHz) occurs over a wide range of physical conditions with kinetic temperatures of T$_k$ = 200 -- 2000 K, hydrogen densities of n(H$_2$) = 10$^{8}$ -- 10$^{10}$ cm$^{-3}$, and \ho densities of n(H$_2$O) = 10$^3$ -- 10$^5$ cm$^{-3}$\citep[e.g.,][]{eli89, neu95, liz05}}. Extragalactic 22\,GHz \ho masers provide important information on
the geometry and kinematics of the central few parsecs of active galactic nuclei (AGN) in terms of dense molecular gas.
{Submillimeter \ho maser transitions at the 321.226\,GHz (10$_{29}$$-$9$_{36}$) and 325.153\,GHz (5$_{15}$$-$4$_{22}$) transitions are  strongly inverted under more restricted physical conditions} (e.g., T$_k$ $>$ 1000 K) \citep{deg77,neu90,neu91}. The first detection of a 321~GHz (10$_{29}$$-$9$_{36}$) \ho maser in a Galactic star-forming region was achieved by \citet{men90}.
Subsequent detections of (sub)millimeter \ho masers at 183.308, 321.226, and 325.153\,GHz in Galactic sources \citep{men90, men91, liz07} have inspired similar searches toward active galaxies.

The studies of extragalactic sub-millimeter \ho masers were pioneered by \citet{liz05},
who discovered extragalactic sub-millimeter \ho masers at the 183\,GHz (3$_{13}$$-$2$_{20}$) and 439\,GHz (6$_{43}$-5$_{50}$) transitions toward the nuclear region of the Type 2 Seyfert/LINER galaxy NGC 3079 using the Submillimeter Array (SMA) and James Clerk Maxwell Telescope (JCMT).
{The detection of 183\,GHz \ho emission from the merging galaxy \object{Arp220}, which hosts two nuclei but no active nucleus in its center, has also been reported \citep{cer06}.}
\citet{hagi13} reported the first extragalactic detection of 321\,GHz \ho maser emission toward the center of the Circinus galaxy, which has a Type 2 Seyfert nucleus, using the Atacama Large Millimeter/submillimeter Array (ALMA). 

These studies suggest that extragalactic sub-millimeter \ho masers may be common in AGN and could be associated with AGN activity or trace a circumnuclear disk around the AGN like the nuclear \ho megamasers at 22\,GHz \citep[e.g.,][]{miyo95}, {although the 183 GHz \ho emission in \object{Arp220} arises from a star-forming site in the galaxy rather than AGN activity \citep{cer06}.}
Following our previous study of 321\,GHz maser emission in the Circinus galaxy, we present here the results of a search for 321\,GHz and 325\,GHz \ho masers toward several AGN, conducted with ALMA.
Throughout this article, cosmological parameters of H$_{0}$ = 73 \kmss  Mpc$
{-1}$, $\Omega$$_{\Lambda}$ = 0.73, and $\Omega$$_{M}$ = 0.27 are adopted.
The optical velocity definition is adopted throughout this article and the velocities are calculated with respect to the Local Standard of Rest (LSR).  
\section{SAMPLE SELECTION}
In this program, we concentrated our search for extragalactic submillimeter \ho masers 
at the 321.226\,GHz transition from the known \ho nuclear masers with the strongest maser flux densities at 22~GHz \citep{cla86,lincgw97,dos79,bra03}, and additionally searched for the 325.153\,GHz transition from the megamaser galaxy, NGC 5793 \citep{hagi97}. 
It is generally understood that 321\,GHz and 325\,GHz \ho masers exist in circumstellar envelopes in Galactic star-forming regions \citep[e.g.,][]{men90,men91}, while the 325\,GHz \ho emission is known to be strongly attenuated by water vapor absorption in atmosphere.
However, the attenuation of the 325\,GHz \ho line is less significant for distant galaxies due to the redshift of the \ho line velocity, and so we included NGC~5793, at distance of 51\,Mpc, in our list (Table~$\ref{table1}$).
All the galaxies in our list contain 22\,GHz nuclear masers {which} are sub-categorized as disk masers, having characteristic spectra implying the presence of a disk around a nucleus \citep[e.g.,][]{linc09}. The presence of a nuclear edge-on disk in the nuclear region  {makes} the detections of submillimeter \ho maser more likely \citep[e.g.,][]{liz05}. {However, it is also possible that submillimeter maser emission
may arise from extragalactic star-forming sites like the 22\,GHz \ho masers in star-forming galaxies \citep[e.g.,][]{linc93, hagi07}.}
\section{OBSERVATIONS AND DATA REDUCTION}
We conducted spectral-line observations at the 321\,GHz and 325\,GHz transitions toward five galaxies in ALMA Band-7 during the Cycle-0 time between 2012 June 2 and June 6 under the experiment code \#2011.0.00121.S (PI: Hagiwara).
\citet{hagi13} have previously reported the detection of 321\,GHz \ho maser {emission} in the nearby Circinus galaxy arising from this set of observations,
and preliminary results from this fuller study were presented by \citet{hagi15}.
A summary of the observations is given in Table~$\ref{table1}$, including the target galaxies, observation dates, and the number of antennas employed in each observation.
We set the detection threshold at the 5$\sigma$~level, with the observed 1$\sigma$ noise level of $\sim$20 mJy resulting in a flux-density limit of about 100 \mb\ for all sources. This was considered to be sufficient to search for new submillimeter \ho masers from the known 22\,GHz megamaser galaxies having a disk or disk-like structure.
The durations of the observations ranged from 20 to 50 minutes (approximately 5--20 minutes on-source), depending on the target source. 
In our all observations, a single dual polarization spectral window (1.875\,GHz bandwidth)
in the Frequency Division Mode was employed: The single 1.875\,GHz bandwidth was divided into 3840 spectral channels, yielding spectral resolutions of 488 kHz or $\sim$ 0.45 \kms at the observed frequency of 319.4--321.5\,GHz.
The resultant total velocity coverage was $\sim$ 1760 \kmss, centered near each galaxy's systemic velocity.
We conducted phase-referencing observations using a bright phase-reference source
within $\sim$10$^\circ$ of the target sources.
Data calibration was performed using the Common Astronomy
Software Applications (CASA).
Amplitude calibration was performed using observations of Titan or Uranus, 
and the bandpass calibration was made using the bright quasar, either 3C\,279 or 3C\,454.3, that was nearest 
to the target source. Flux-density errors of 10 percent are adopted, as stated in
the capabilities of Cycle 0 observations for Band 7.
The image analysis was performed using both CASA and the Astronomical Image Processing Software (AIPS).
After phase and amplitude calibration, the 321/325\,GHz continuum emission was subtracted from the spectral line visibilities using line-free channels prior to the imaging and CLEAN deconvolution. The line emission from each galaxy was separated out from the continuum emission. The synthesized beam sizes (natural-weighting) and the beam position angles used in the CLEAN process and the resultant rms noise levels in the spectral line images are listed in Table~$\ref{table1}$.
%
\section{RESULTS}
\begin{figure}[ht!]
\plotone{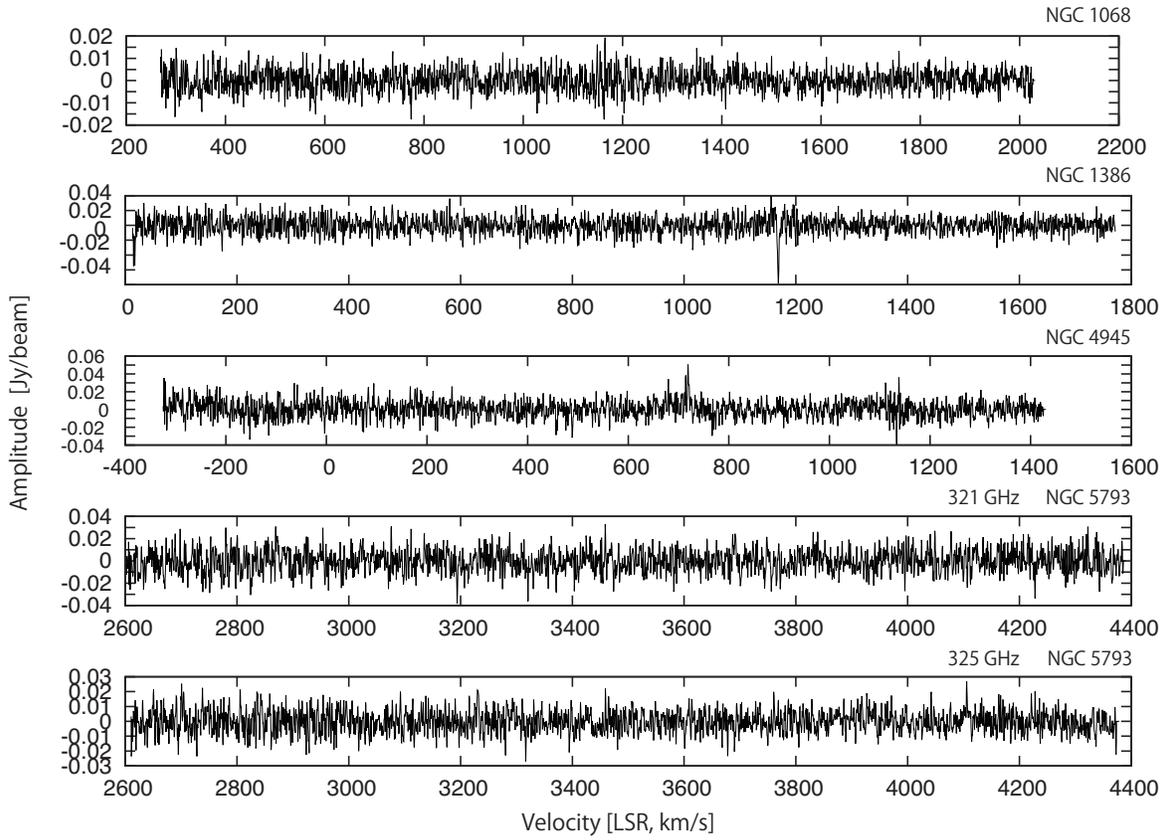}
\caption{Spectra of NGC\,1068, NGC\,1386, NGC\,4945, and NGC\,5793 at 321\,GHz and a spectrum of NGC\,5793 at 325\,GHz, obtained by the ALMA between 2012 June 2--6. All of these spectra were obtained toward the peak of the submillimeter continuum emission centered at each observing frequency. The negative components seen in the NGC\,1386 spectra are not real, {and arise from insufficient amplitude calibration for those channels}.
\label{fig1}}
\end{figure}

\begin{figure}
%
\plotone{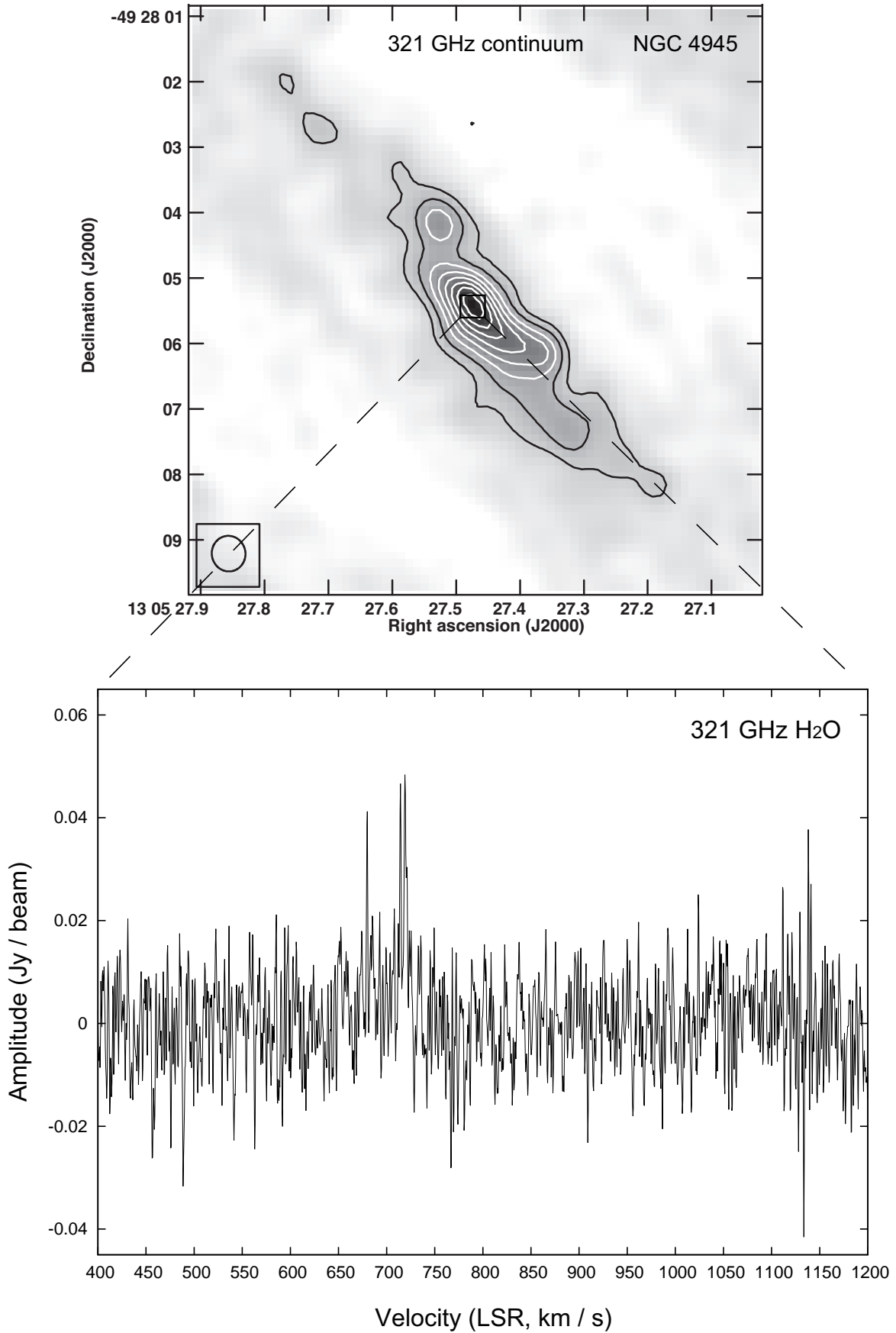}
\caption{321\,GHz continuum image and {the spectrum of \ho maser} in NGC\,4945 toward the center of the continuum, obtained by ALMA on 2012 June 3.
Intensity-scales are denoted by contours and grey-scales: The contour levels
are $-$3, 3, 5, 7, 9, 11, 13, 15, 17, 19 of 7.35 \mb ($\sim$ 1 $\sigma$), the peak flux
density is 136.5 \mb, and the grey-scale ranges from 10 to 136.0 \mb.
The synthesized beam is shown in the bottom left. The total velocity coverage of the spectrum is \vlsr=400--1200 \kmss. The maser spectrum is obtained toward the region marked by a box on the continuum image.  
\label{fig2}}
\end{figure}

We searched for maser emission in each AGN within a field of view of $\sim$ 18\arcsec, centered on the phase-tracking center of each source and within the observed velocity range.
Fig.~\ref{fig1} shows spectra at 321\,GHz from the four AGN, 
NGC\,1068, NGC\,1386, NGC\,4945, and NGC\,5793, and at 325\,GHz from NGC\,5793.
The only emission exceeding a signal-to-noise ratio (SNR) of 5 is seen in NGC\,4945, as reported by \citet{hagi15}.
The detection of \ho emission toward NGC\,4945
is reminiscent of that towards the Circinus galaxy reported in our earlier study
\citep{hagi13}. {Given the narrow line-widths of the \ho emission and the fact that it is unresolved at the angular resolution of our observations, it is most likely that the detected \ho emission has a maser origin, as in the case of Circinus.}
Fig.~\ref{fig2} displays {the spectrum of the \ho maser emission} and the 321\,GHz continuum emission in NGC\,4945. Several 321\,GHz \ho maser features were detected in NGC\,4945, {with} peak flux densities of {$\sim$ 45 \mb~at \vlsr = 714.4 \kms} and $\sim$ 50 \mb~ at \vlsr = 718.9 \kmss: All these features are detected with a SNR $>$ 5. 
In Fig.~\ref{fig1}, the detected \ho emission of NGC\,4945 comprises distinct features that are redshifted from the systemic velocity of the galaxy (\vlsr=556 \kmss). The feature at \vlsr=714.4 \kms, which shows the strongest peak, is redshifted by 162.9 \kms from the systemic velocity. Most of the other prominent features lie in the redshifted velocity range of \vlsr=711--722 \kms, with the exception of a narrow feature at \vlsr=679.9 \kms with a peak flux density of {$\sim$40 \mb}. The total integrated intensity estimated from these features lying at \vlsr=  679--680 \kms and  \vlsr=712--722 \kms is $\sim$ 2.6 Jy \kmss, which corresponds to $\sim$ 7.4 \lsun.
We note a highly redshifted feature at the  
$\sim$3$\sigma$ level centered at \vlsr = 1138.1 \kms with a peak strength $\sim$35 \mb (Fig.~\ref{fig1},~\ref{fig2}).
{The corresponding fringe phases do not clearly show a consistent trend.}
Further, more sensitive observations are required to confirm this feature.

\begin{figure}
\plotone{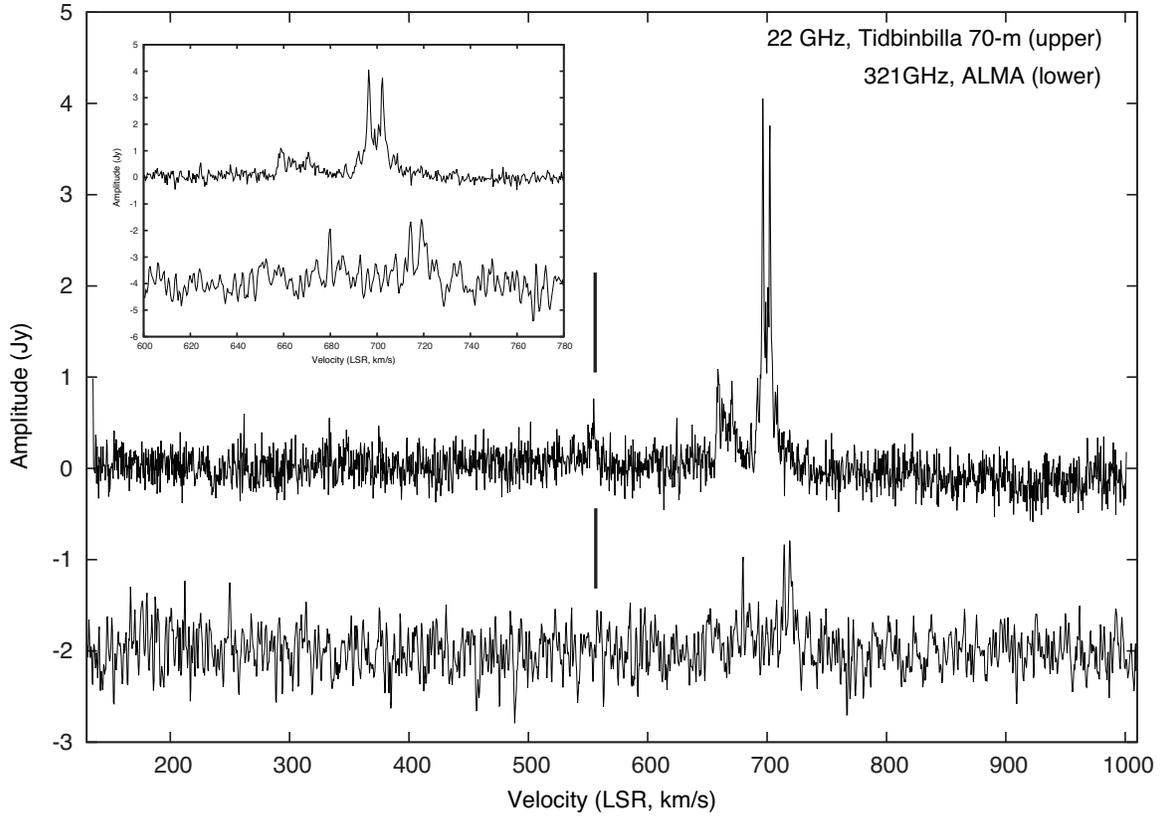}
\caption{Comparison of \ho maser spectra in the 321\,GHz and 22\,GHz transitions toward the continuum nucleus of NGC\,4945 in velocities spanning from \vlsr=130 \kms to \vlsr=1100~\kmss. The 22\,GHz spectrum was obtained with the Deep Space Network (DSN) 70-m antenna at Tidbinbilla on 2014 April 7. The amplitude scale of the 321\,GHz spectrum is multiplied by a factor of 25 
 {with an offset of 2 Jy for display purposes}. {The systemic velocity of NGC\,4945 is denoted by vertical bars}. The inset displays the two maser spectra in velocities from \vlsr=600--780 \kmss, where the amplitude scale of the 321\,GHz spectrum is multiplied by a factor of 50
and an offset of 4 Jy applied for clarity. \label{fig3}}
\end{figure}

In Fig.~\ref{fig3}, we compare the 321\,GHz \ho spectrum with a 22\,GHz spectrum obtained on
2014 April 7 using the Deep Space Network 70-m antenna at Tidbinbilla. This observation was conducted at a spectral resolution of 31.25 kHz or 0.42 \kms and with an rms of $\sim$0.3 Jy. We conservatively estimate that the uncertainty in amplitude is up to 50 percent.
The known strong variability of the 22\,GHz maser {emission} in this galaxy, together with the $\sim$1.8 years between the two observations, make it difficult to precisely compare the velocities of each feature between the 22\,GHz and 321\,GHz lines. In our 321\,GHz observation, no feature was detected at or near the systemic velocity, while the 22\,GHz {spectrum} in Fig.~\ref{fig3} shows a systemic feature, which is consistent with earlier single-dish measurements,  \citep[e.g.,][]{linc97}. 
It should be noted that an asymmetry between the blueshifted and redshifted emission is commonly observed in \ho megamasers, such as NGC 4258: Redshifted features are more numerous and are significantly more intense than the blueshifted features \citep[e.g.,][]{mao98}.

A comparison of the 321\,GHz \ho spectrum with the 22\,GHz spectrum in Fig.~\ref{fig3}, and similar spectra in the literature \citep{naka95,bra03}, indicates that the detected 321\,GHz emission in the galaxy is similar to the features appearing in the 22\,GHz maser spectra spanning \vlsr = 643--714 \kmss, however, there is no direct correspondence between the velocities of the features in the two bands.
In addition, in the published 22\,GHz maser spectrum of \citet{linc97}, the most redshifted velocity feature at \vlsr = 774.4 \kms was tentatively detected, whereas there is no corresponding feature
in either our 321\,GHz or 22\,GHz spectra.
(Velocity values are converted from optical heliocentric to the LSR definition, using the relation \vlsr = V$_{hel}$ -- 4.6 \kms in \citet{linc97}). 
%
%
Based on the ALMA data, the Gaussian-fitted phase-referenced position of the \vlsr = 718.0 \kms peak emission is $\alpha$(2000):13$^h$05$^{\rm m}$27$\fs$48, $\delta$(2000): $-$49$\degr$28$\arcmin$ 05$\arcsec$.50. Positional uncertainties of 0$\arcsec$.03 are estimated.
The position of the marginally significant redshifted emission at \vlsr = 1138.6 \kms coincides with this peak within the uncertainties.  According to the ALMA Cycle 0 capabilities, the positional errors 
are $\sim$ 10\% of the synthesized beam, corresponding to 0$\arcsec$.54/10 $\sim$ 0$\arcsec$.054 in our observation, and this can be taken as the most dominant source of error.
The relative positions of other redshifted features coincide within uncertainties and the features remain unresolved at the angular resolution of $\sim$ 0$\arcsec$.54, or $\sim$30pc.
The position of the 22\,GHz \ho maser ($\alpha$(2000):13$^h$05$^{\rm m}$27$\fs$48$\pm$0.02, $\delta$(2000): $-$49$\degr28\arcmin 05\arcsec$.4 $\pm$ 0.1; \citet{linc97}) coincides with those of the 321\,GHz maser within uncertainties.

A 321\,GHz submillimeter continuum image is obtained toward the nucleus of the galaxy, with a Gaussian-fitted peak position of $\alpha$(2000):13$^h$05$^{\rm m}$27.47, $\delta$(2000): $-$49$\degr28\arcmin 05\arcsec.42$ (with uncertainties of $\sim$0$\arcsec$.015). The continuum peak flux density at this resolution is 136.5 \mb, which is detected with an SNR of $\sim$20 (Fig.~\ref{fig2}) and an rms noise of about 7\,\mb, with a total continuum flux density of 543.3\,\mb.
Thus, the approximate relative positional errors between the \ho emission and the continuum peak, represented by the synthesized beam size divided by 2 $\times$ SNR \citep{hagi01}
are $\sim$0$\arcsec$.055 or 3.0\,pc, within which the maser resides in the submillimeter continuum nucleus. Therefore, we conclude that the locations of the 321\,GHz \ho maser spots coincide with that of the submillimeter continuum peak within uncertainties of $\sim$3.0 pc.
%
%
%
\section{DISCUSSION}
\subsection{The 321 GHz continuum and new maser emission in NGC\,4945}
It is clear that the 321\,GHz \ho maser {emission} was detected toward the center of NGC\,4945, an edge-on spiral housing a Type 2 Seyfert nucleus \citep{har85}. The known, strong 22\,GHz \ho nuclear maser {emission} in this galaxy \citep{dos79} exhibits a disk-like structure around the nucleus \citep{linc97}. X-ray observations revealed that 
the nucleus of the galaxy is heavily obscured by foreground material
\citep{iwa93,iso08},
which is consistent with the fact that the maser excitation occurs in a region where {there is abundant molecular material} along our line of sight. 
The 321\,GHz continuum emission shows a structure that is partly resolved in the northeast direction accompanying an elongated substructure in the north (Fig.~\ref{fig2}).
Given the synthesized beam size and position angle in Fig.\ref{fig2}, the major axis of the structure is real and similar to that of the CO galactic disk of $\sim$45$\degr$ $\pm$ 2$\degr$ \citep{dah93}.
The Spectral Energy Distribution of the galaxy (primarily from NED) suggests that the submillimeter continuum of the galaxy is dominated by dust emission: the photometric data points show that the continuum flux peaks at submillimeter wavelengths. Thus, it is likely that the most dominant submillimeter continuum component traces the dust lanes lying over the galactic disk, and
that are heated by an AGN in the obscured nucleus,
%
As discussed in the previous section, the 321\,GHz \ho maser {emission coincides with} the nuclear continuum peak, and the maser is likely to be associated with AGN-activity in the galaxy by analogy with the case of 22\,GHz maser in the galaxy. However, we cannot rule out {the possibility} that the maser {emission} originates in star-forming activity or nuclear starbursts. 
The angular resolution of our Cycle-0 observations were not sufficient to resolve these maser spots
and clarify the origin of the maser emission.
The 22\,GHz and 183\,GHz \ho emission in the LINER/Seyfert Type 2 galaxy, NGC\,3079
occur in a similar velocity range 
\citep{liz05}. This suggests that these masers are located in similar regions, at least, in our line of sight. As with the case of the masers in the Circinus galaxy discussed in \cite{hagi13}, the difference of locations of the 321\,GHz and 22\,GHz \ho maser must be addressed also for NGC\,4945.

Simultaneous monitoring of these 22\,GHz and 321\,GHz transitions that are ortho-\ho would help to clarify the excitation mechanism of the masers, by constraining radiative transfer models \citep[e.g.][]{liz07}.
\subsection{Submillimeter \ho masers in AGN}
Presently, only a handful of extragalactic submillimeter \ho masers have been detected. In this observing program, submillimeter \ho masers were discovered {in two of} the five AGNs. The detections are from two nearby AGN hosting {the} 22\,GHz nuclear masers with the strongest extragalactic maser flux densities ($\ga$10 Jy) known to date \citep{dos79,linc03,hagi13}. The detection of the submillimeter \ho masers can therefore be predicted for the strongest nuclear masers known at 22\,GHz in the northern sky, primarily in the nearby AGN, NGC\,4258 and possibly in NGC\,3079, although the latter is difficult to observe with ALMA. The non-detections for the other AGN in our list might not be surprising, as their 22\,GHz flux densities are an order of magnitude smaller than those from strongest nuclear masers.

We did not detect maser emission at either frequency for NGC\,5793. It is known that 325\,GHz \ho masers in star-forming regions are strongly inverted like 22\,GHz masers\citep{men90b}. The 22\,GHz maser in this galaxy is not as bright as the other strong nuclear masers, and so the non-detection of the maser is not surprising.

In previous studies of \ho masers in star-forming regions, the total velocity spread of 321\,GHz \ho masers is observed to be smaller than that of 22\,GHz \ho maser \citep{men90}, while our data marginally detected highly redshifted features (redshift up to $\sim$600 \kmss) in both the Circinus galaxy and NGC\,4945, from which we can speculate that extragalactic submillimeter masers are tracing phenomena which {do not originate from star-forming activity} but {rather from AGN activity}.
The
variability of the submillimeter \ho maser {emission} in the Circinus galaxy was discussed in \citet{hagi13}, 
who inferred that the 22\,GHz maser was in a flaring state in mid-2012; significant increases of the maser flux density up to 50--70 Jy were measured in 2012 June and September (note that the flux values include large uncertainties). This is the largest flare ever reported in this galaxy \cite[e.g.,][]{linc97,bra03}. Our detection of the 321\,GHz \ho maser in the Circinus {galaxy}  might be due to the large flare in 2012. Likewise, the detection threshold of submillimeter \ho maser may depend strongly on the flux variability.
%
%
\subsection{High-Velocity Dense Gas in NGC\,4945}
\citet{linc97} estimated the central binding mass in NGC\,4945, by assuming the rotation of gases on a Keplerian disk of 150 \kms at 0.3 pc from the central engine and a disk inclination of 90$\degr$, which yielded a black hole mass of 1.4 $\times$ 10$^6$ \sm. 
In our observations, a highly redshifted velocity feature at \vlsr= 1138.6 \kms is marginally detected.
In their analysis of this data set, \citet{pes16} note that this velocity range is impacted by higher atmospheric absorption resulting in a somewhat higher noise level, and so further observations are required to confirm this feature.

We note, however, that if the 1138 \kms feature, redshifted by  $\sim$585 \kms from the systemic velocity, is real, the maser is located at a distance of $\sim$0.020 pc from the central engine, based on {the} Keplerian ($v$ $\propto$ $r^{-0.5}$) rotating disk model in \citet{linc97}. This corresponds to $\sim$ 1.5 $\times$ 10$^5$ Schwarzschild radii for a 1.4 $\times$ 10$^6$ \sm black hole.
Similarly, the most redshifted velocity feature 
in the Circinus (redshifted by $\sim$ 635 \kms from the systemic) is estimated to be
at a radius of $\sim$0.018 pc ($\sim$ 1.2 $\times$ 10$^5$ Schwarzschild radii for a 1.7 $\times$ 10$^6$ $\sm$ black hole) \citep{hagi13}.
One can speculate that the high-velocity dense gas found in other AGNs in future observations is tracing the molecular material closest to central engine.
{High-resolution imaging of submillimeter masers promises to be a powerful tool
to further explore the gas that probes the AGN circumnuclear region on parsec- and sub-parsec scales}.
Since the masers detected in this program are unresolved, their spatial structures have yet to be explored for studying the circumnuclear regions of their host galaxy. {ALMA has progressively achieved longer baselines and has attained a maximum angular resolution of $\sim$30 mas (corresponding to 1.5\,pc at a distance of 10\,Mpc) in band 7 \citep[e.g.,][]{br15, ed15}}. Future observations will be able to resolve the circumnuclear gas of AGN on scales that are becoming comparable to VLBI observations at 22\,GHz. 
\acknowledgments
This research was supported by Japan Society for the Promotion of Science (JSPS) Grant-in-Aid for Scientific Research (B) (Grant Number: 15H03644).
This article makes use of the following ALMA data: \#2011.0.00121.S. ALMA is a partnership of ESO (representing its member states), NSF (USA) and NINS (Japan), together with NRC (Canada) and NSC and ASIAA (Taiwan), in cooperation with the Republic of Chile. The Joint ALMA Observatory is operated by ESO, AUI/NRAO and NAOJ.  This research has made use of the NASA/IPAC Extragalactic Database (NED) which is operated by the Jet Propulsion Laboratory, California Institute of Technology, under contract with the National Aeronautics and Space Administration. The Tidbinbilla 70-m telescope is part of the NASA Deep Space Network and is operated by CSIRO. 
\vspace{5mm}
\facilities{ALMA, Tidbinbilla 70-m telescope}
\floattable
%
\begin{deluxetable}{lccccccccc}
\tabletypesize{\scriptsize}
\tablecaption{Summary of a search for extragalactic submillimeter \ho emission  \label{table1}}
\tablewidth{0pt}
\tablehead{\colhead{Source}&\colhead{RA$^{\rm a}$}&\colhead{DEC$^{\rm a}$}&D$^{\rm b}$&\colhead{Date$^{\rm c}$}&\colhead{$\nu$$^{\rm d}$}&\colhead{N$_{\rm A}$$^{\rm e}$}& \colhead{t$_{\rm on}$$^{\rm f}$}
&\colhead{$\theta_{b}$$^g$ (PA$^{\rm h}$)}&\colhead{1~$\sigma$$^{\rm i}$}\\
&\colhead{(J2000)}&\colhead{(J2000)}&(Mpc)&&\colhead{(GHz)}&&(min)
&\colhead{(arcsec$^2$, $^{\circ}$)}&\colhead{(mJy)}
} 
\startdata
NGC\,1068 & 02$^{\rm h}$42$^{\rm m}$40$^{\rm s}$.770 & $-$00$^{\circ}$00${\arcmin}$47${\arcsec}$.84 & 12.5 & June 2-6 & 319.07 & 21 & 39 & 0.61$\times$0.44 (42)     &   4--6 \\
NGC\,1386 & 03$^{\rm h}$36$^{\rm m}$46$^{\rm s}$.237 & $-$35$^{\circ}$59${\arcmin}$57${\arcsec}$.39 & 10.6 & June 5 & 319.36& 29 & 11 & 0.67$\times$0.54 ($-$27)  & 10--15 \\
NGC\,4945 & 13$^{\rm h}$05$^{\rm m}$27$^{\rm s}$.279 & $-$49$^{\circ}$28${\arcmin}$04${\arcsec}$.44 & 11.1 & June 3 & 319.68 & 18 & 14 & 0.54$\times$0.51 (11)     &  8--11 \\
Circinus  & 14$^{\rm h}$13$^{\rm m}$09$^{\rm s}$.906 & $-$65$^{\circ}$20${\arcmin}$20${\arcsec}$.46 &  4.2 & June 3 & 319.82 & 18 & 18 & 0.66$\times$0.51 ($-$17)  &  9--11 \\
NGC\,5793 & 14$^{\rm h}$59$^{\rm m}$24$^{\rm s}$.807 & $-$16$^{\circ}$41${\arcmin}$36${\arcsec}$.55 & 51.1 & June 3 & 316.59 & 21 &  6 & 0.56$\times$0.47 (47)     & 11--13 \\
NGC\,5793 & 14$^{\rm h}$59$^{\rm m}$24$^{\rm s}$.807 & $-$16$^{\circ}$41${\arcmin}$36${\arcsec}$.55 & 51.1 & June 3 & 320.47 & 21 & 19 & 0.66$\times$0.46 ($-$89)  &  8--11 \\
\enddata
\tablenotetext{a}{~Interferometry phase center positions in R.A.\ and Declination}
\tablenotetext{b}{~Luminosity distances, adopted from NED}
\tablenotetext{c}{~Date of observations in 2012}
\tablenotetext{d}{~Observing frequencies at channel 0 in each spectral window}
\tablenotetext{e}{~Number of the 12-m antenna used in the ALMA observations}
\tablenotetext{f}{~On-source time for the target source}
\tablenotetext{g}{~Synthesized beam size}
\tablenotetext{h}{~The beam position angle}
\tablenotetext{i}{~rms noise values in a 488.3 kHz spectral resolution in the clean image, depending on channel}
\end{deluxetable}
%
\clearpage

\end{document}